%% file: article.tex
\documentclass[pdflatex,sn-mathphys-num]{sn-jnl}

\usepackage{graphicx}%
\usepackage{multirow}%
\usepackage{amsmath,amssymb,amsfonts}%
\usepackage{amsthm}%
\usepackage{mathrsfs}%
\usepackage[title]{appendix}%
\usepackage{xcolor}%
\usepackage{textcomp}%
\usepackage{manyfoot}%
\usepackage{booktabs}%
\usepackage{algorithm}%
\usepackage{algorithmicx}%
\usepackage{algpseudocode}%
\usepackage{listings}%

\usepackage[utf8]{inputenc} 
\usepackage[T1]{fontenc}    
\usepackage{hyperref}       
\usepackage{url}            
\usepackage{booktabs}       
\usepackage{amsfonts}       
\usepackage{nicefrac}       
\usepackage{microtype}      
\usepackage{xcolor}         

\usepackage{amsmath}
\usepackage{graphicx}
\usepackage{algorithm}
\usepackage{algpseudocode}
\usepackage{amsmath}
\usepackage{adjustbox}
\usepackage{graphicx}
\usepackage{caption}

\usepackage{amsmath, amssymb}
\usepackage{caption}
\usepackage{subcaption}
\usepackage{multicol}

\usepackage{algpseudocode}
\usepackage{amsmath}
\usepackage{amssymb}
\usepackage{geometry}
\theoremstyle{thmstyleone}%
\theoremstyle{thmstyletwo}%

\theoremstyle{thmstylethree}%

\usepackage{amsmath}

\begin{document}
\title[Article Title]{Cryo-SWAN: the Multi-Scale Wavelet-decomposition-inspired Autoencoder Network for molecular density representation of molecular volumes}



\author[1,2]{\fnm{Rui} \sur{Li}}

\author[3,4]{\fnm{Artsemi} \sur{Yushkevich}}

\author[3,5]{\fnm{Mikhail} \sur{Kudryashev}}\email{mikhail.kudryashev@mdc-berlin.de}

\author[1,2,6,7]{\fnm{Artur} \sur{Yakimovich}}\email{a.yakimovich@hzdr.de}

\affil[1]{\orgdiv{Center for Advanced Systems Understanding (CASUS)}, \orgaddress{\city{Görlitz}, \country{Germany}}}

\affil[2]{\orgdiv{Helmholtz-Zentrum Dresden-Rossendorf e. V. (HZDR)}, \orgaddress{\city{Dresden}, \country{Germany}}}

\affil[3]{\orgdiv{In situ Structural Biology}, \orgname{Max Delbrück Center for Molecular Medicine in the Helmholtz Association}, \orgaddress{\city{Berlin}, \country{Germany}}}

\affil[4]{\orgdiv{Department of Physics}, \orgname{Humboldt University of Berlin}, \orgaddress{\city{Berlin}, \country{Germany}}}

\affil[5]{\orgdiv{Institute of Medical Physics and Biophysics}, \orgname{Charite-Universitätsmedizin}, \orgaddress{\city{Berlin}, \country{Germany}}}

\affil[6]{\orgdiv{Institute of Computer Science}, \orgname{University of Wrocław}, \orgaddress{\city{Wrocław}, \country{Poland}}}

\affil[7]{\orgdiv{Cluster of Excellence Physics of Life}, \orgname{TU Dresden}, \orgaddress{\city{Dresden}, \country{Germany}}}


\abstract{
Learning robust representations of 3D shapes from voxelized data is essential for advancing AI methods in biomedical imaging. However, most contemporary 3D computer vision approaches operate on point clouds, meshes, or octrees, while volumetric density maps, the native format of structural biology and cryo-EM, remain comparatively underexplored. We present Cryo-SWAN, a voxel-based variational autoencoder inspired by multi-scale wavelet decomposition. The model performs conditional coarse-to-fine latent encoding and recursive residual quantization across perception scales, enabling accurate capture of both global geometry and high-frequency structural detail in molecular density volumes. Evaluated on ModelNet40, BuildingNet, and a newly curated dataset of cryo-EM volumes, ProteinNet3D, Cryo-SWAN consistently improves reconstruction quality over state-of-the-art 3D autoencoders. We demonstrate that the molecular densities organize in learned latent space according to shared geometric features, while integration with diffusion models enables denoising and conditional shape generation. Together, Cryo-SWAN is a practical framework for data-driven structural biology and volumetric imaging.
}


\maketitle

\section*{Introduction}\label{intro}
\input{sec/1_intro}

\section*{Methods}\label{methods}
\input{sec/4_methods}
\section*{Results}\label{res}

\input{sec/2_results}

\section*{Discussion}\label{discussion}
\input{sec/3_discussion}


\section*{Acknowledgments}
\input{sec/5_acknowlegements}


\bibliography{ref}

\section*{Supplementary Information}
\input{sec/6_suppl}

\end{document}

%% file: sec/1_intro.tex
The shape of 3D objects is often directly connected to their function. Learning a meaningful representation of objects' shapes from noisy input data is paramount to accomplishing advanced machine learning tasks downstream. Such tasks could facilitate advancements in scientific fields, including cellular and structural biology, aiming to connect the shape of organelles and molecules to their biological functions. Through this, structural biology facilitates the understanding of important biological mechanisms and ultimately therapies. 

In biomedical imaging, 3D shapes are typically stored in voxel space. However, in 3D computer vision, most methods operate on processed geometric representations such as octrees \cite{xiong_octfusion_2024} or VDB grids \cite{museth_vdb_2013}. Approaches such as Craftsman \cite{li_craftsman_2024}, XCube \cite{ren_xcube_2024}, and Dora-VAE \cite{chen_dora_2025} exploit point clouds, sparse voxels, or meshes. In parallel to the biomedical imaging, the task of 3D shape generation \cite{ren_xcube_2024, chen_dora_2025, cheng_sdfusion_2023, mittal_autosdf_2023, li_craftsman_2024, zhao_michelangelo_2023, li_advances_2024} has gained increasing attention due to rapid progress in AR/VR, animation, and autonomous driving. This led to the voxel-based representation learning being largely underexplored. Meaningful representations are also crucial for the tasks involving modern generative models guided by learned priors \cite{blattmann_align_2023, lee_autoregressive_2022, razavi_vqvae2_2019, rombach_high-resolution_2022, takida_hq-vae_2024, tian_visual_2024}. Once a desired shape can be generated it opens versatile opportunities in drug design and \textit{in silico} screening.

Most state-of-the-art (SOTA) 3D generative frameworks adopt a two-stage paradigm. First, they learn expressive latent representations using an autoencoder (AE) \cite{kingma_auto-encoding_2014}. Second, they train generative models, such as diffusion models \cite{ho_denoising_2020}, generative adversarial networks (GANs) \cite{goodfellow_generative_2014}, or transformers \cite{vaswani2017attention} on the learned latent space. A typical AE \cite{kingma_auto-encoding_2014,bank2023autoencoders} consists of an encoder, a latent layer, and a decoder. The encoder maps a high-dimensional input $x_i \in \mathcal{X}$ to a latent vector $z_i \in \mathcal{Z}$ in a process known as representation learning \cite{bengio2013representation}, and a decoder reconstructs it as $x_i' \in \mathcal{X'}$. The quality of the learned representation determines downstream generative performance, especially for high-dimensional and geometry-rich data such as 3D shapes. Most AE-based approaches in structural biology rely on either 2D projections or atomic models. Yet, cryo-EM data, stored as 3D density volumes, rather than atomic structures, leaves representation learning on 3D cryo-EM densities underexplored.

The AE-based reconstruction process is governed by an encoding distribution $q(\mathcal{Z} \mid \mathcal{X})$ and a decoding distribution $p(\mathcal{X'} \mid \mathcal{Z})$. The conventional AEs optimize only the reconstruction loss $|x_i' - x_i|_2$ and do not constrain the latent structure. Variational autoencoders (VAEs), on the other hand, introduced a Gaussian prior over the latent space \cite{kingma_auto-encoding_2014}, enabling sampling and regularization via the KL-divergence \cite{odaibo_tutorial_2019}. The latent variables are constrained as ( $z_i \sim \mathcal{N}(0,1)$ ), and training includes a regularization term ( $\beta D_{\text{KL}}(q_\phi(z|x) || p(z))$ ). Although VAEs improve latent organization and training stability, they often produce blurry reconstructions due to overly coarse embeddings \cite{kossale_mode_2022, zeghidour_soundstream_2022}. As a result, recent research has focused on improving latent expressiveness. VQ-VAE \cite{oord_neural_2018} replaced continuous latent variables with a discrete codebook $e \in \mathbb{R}^{K \times D}$. Given an encoder output $z_e(x)$, discretization is performed via nearest-neighbor lookup:
\[
q(z = k \mid x) =
\begin{cases}
1 & \text{if } k = \arg\min_j | z_e(x) - e_j |_2, \cr
0 & \text{otherwise}.
\end{cases}
\]
Training includes a reconstruction loss and a commitment loss,
\[
\left| \text{sg}[z_e(x)] - e \right|_2^2 + \beta \left| z_e(x) - \text{sg}[e] \right|_2^2,
\]
where $\text{sg}[\cdot]$ denotes the stop-gradient operator.

The authors of the RQ-VAE introduced recursive residual quantization \cite{lee_autoregressive_2022}. Starting with the encoder output $r^{(0)} = z_e$, the remaining residual errors are quantized across $L$ recursive levels using codebooks $\mathcal{C}^{(l)} = { e_k^{(l)} }$. At each residual level,
\[
c^{(l)} = \arg\min_k | r^{(l-1)} - e_k^{(l)} |_2, \quad
r^{(l)} = r^{(l-1)} - e_{c^{(l)}}^{(l)},
\]
and the final latent approximation is $\hat{z}_L = \sum_{l=1}^L e_{c^{(l)}}^{(l)}$.

HQ-VAE \cite{takida_hq-vae_2024} introduced hierarchical codebooks across multiple spatial scales. The encoder feature map is discretized into latent variables $\mathbf{Z}_{1:L} = { \mathbf{Z}^{(l)} }_{l=1}^L$, where each spatial scale level has a dedicated codebook $\mathcal{C}^{(l)} = { e_k^{(l)} }$. Commitment losses are computed hierarchically:
\[
\mathcal{L}_{\text{code}} = \sum_{l=1}^{L} \sum_{i=1}^{d_l}
\left| \text{sg}[\hat{z}_i^{(l)}] - e_k^{(l)} \right|^2.
\]

VAR-VAE \cite{tian_visual_2024} adopted a multi-scale tokenization strategy inspired by transformer-based vision models \cite{esser_taming_2021, prabhakar2024vit, chefer2021transformer}. Latent maps $\mathbf{Z}^{(l)}$ are embedded at progressively coarser resolutions and modeled autoregressively:
\[
p(\mathbf{Z}^{(1:L)}) = \prod_{l=1}^{L} p\left( \mathbf{Z}^{(l)} \mid \mathbf{Z}^{(<l)} \right).
\]

In this work we introduce cryo-SWAN, a multi-Scale quantized Wavelet-inspired  \cite{li2024solving, fan_bcr-net_2019} AutoeNcoder, capable of effectively learning meaningful shape representations from the voxel space. Cryo-SWAN performs multi-level latent quantization combined with recursive residual optimization at each scale, enabling high-fidelity representation learning directly from raw density volumes without geometric or atomic priors. Multi-scale representations are widely used in image restoration and signal approximation \cite{zamir_multi-stage_2021, zhou2016perceptual, li2024solving}. For example, MPRNet \cite{zamir_multi-stage_2021} and BCR-Net \cite{fan_bcr-net_2019, li2024solving} demonstrated that recursive, multi-level decomposition effectively preserves high-frequency information, which is critical for accurate reconstruction. Here, we demonstrate the applicability of multi-scale representations to 3D shapes.

We evaluate cryo-SWAN on two standard computer vision benchmarks: ModelNet \cite{zhirong_wu_3d_2015} and BuildingNet \cite{selvaraju_buildingnet_2021}. Furthermore, we introduce ProteinNet3D, a new dataset comprising over 24k experimental cryo-EM density maps from EMDB \cite{the_wwpdb_consortium_emdbelectron_2024}, manually curated. Cryo-SWAN is compared to the state-of-the-art voxel-based AE models, including HQ-VAE \cite{takida_hq-vae_2024}, VAR-VAE \cite{tian_visual_2024}, RQ-VAE \cite{lee_autoregressive_2022}, and VQ-VAE \cite{oord_neural_2018}. We report IoU, F1-score, MSE, and PSNR across all datasets, and additionally evaluate reconstruction resolution on ProteinNet3D using Fourier Shell Correlation (FSC) \cite{van2005fourier, aiyer2020evaluating, agard2014single}. Cryo-SWAN consistently outperforms competing methods, particularly in capturing high-frequency structural details critical for accurate 3D shape representation. Finally, we demonstrate two downstream applications: geometry-based hub detection using latent molecular embeddings and conditional latent diffusion for high-fidelity molecular volume generation, highlighting cryo-SWAN’s potential as a general backbone for data-driven structural biology.

%% file: sec/4_methods.tex
\subsection*{Benchmark datasets: ModelNet40 and BuildingNet}
The ModelNet40 dataset includes 40 categories of common 3D objects in \texttt{.off} format, spanning low-frequency shapes (e.g., chairs) to high-frequency structures (e.g., plants with intricate foliage). The BuildingNet dataset contains \texttt{.obj} files of architectural structures. Unlike ModelNet, architectures are hollow, structurally detailed, and rich in high-frequency components. Thus, it poses more challenges. To match the volumetric nature of cryo-EM data, we voxelized the \texttt{.off} and \texttt{.obj} files as densities.

\subsection*{ProteinNet3D dataset}

To evaluate our model, we curated a 3D volumes dataset of diverse macromolecules obtained from the publicly available Electron Microscopy Data Bank (EMDB). EMDB is a comprehensive, annotated repository of volumetric data derived from experimental and computational cryo-EM. It covers a wide range of macromolecules, molecular complexes, and subcellular structures.

We retrieved the complete EMDB metadata as of late November 2022 via the EMDB API, comprising approximately 28,700 molecular density volumes. To ensure consistency and enable meaningful evaluation, we restricted our selection to entries derived from cryo-EM single-particle analysis (SPA) or cryo-electron tomography (cryo-ET) subtomogram averaging (STA). Focusing on individual macromolecules, we further filtered the dataset by molecular weight, retaining entries within the 100--1500 kDa range. This criterion excluded very small structures (e.g., individual domains) as well as excessively large complexes or subcellular assemblies. To ensure uniform scaling of volumetric features, we normalized voxel spacing by resampling all volumes to a common isotropic pixel. Serving for the deep learning training, we enriched the data based on reported volume size, voxel spacing, and resolution from EMDataResource (http://www.emdataresource.org). Since resampling can affect resolution due to the Nyquist limit, we carefully considered potential resolution degradation introduced by the new sampling rate.

Following the filtering procedure, 5,222 qualified EMDB entries were retained. The associated volumetric data were retrieved through the EMDB API using their accession identifiers. We masked the target volumes to suppress background noise and extraneous regions based on the provided contour level annotation on each entry. All volumes were uniformly resampled to an isotropic voxel spacing of 4~\AA{}, allowing for the optimal volume size (Supplementary Fig.~\ref{metadata_apix}, top) and the optimal Nyquist limit relative to the entry resolution and pixel size originally reported (Supplementary Fig.~\ref{metadata_apix}, bottom). To mitigate aliasing artifacts introduced during resampling, a low-pass filter at the Nyquist frequency was applied. The processed volumes were then adjusted to a fixed spatial resolution of $64^3$ voxels via cropping or zero-padding. We then normalized all entries as zero mean and unit variance.

To improve data diversity, we augmented each volume with five randomly generated 3D rotations. This yields a total of 26,110 samples, while each accompanied by its corresponding metadata.

The breadth of the ProteinNet3D dataset is evidenced by its molecular weight distribution (Supplementary Fig.~\ref{metadata_mw}, top), which ranges from 100 to 1500~kDa (Supplementary Fig.~\ref{metadata_mw}, bottom). Additionally, since the dataset originates from experimentally reconstructed cryo-EM volumes, it contains both canonical macromolecular structures (Supplementary Fig.~\ref{complexity}, top) and reconstruction-related artifacts (Supplementary Fig.~\ref{complexity}, middle). In addition, because numerous proteins operate within membrane-associated environments, membrane densities are frequently retained in the reconstructed volumes (Supplementary Fig.~\ref{complexity}, bottom).

\subsection*{Model training hyperparameters}
All datasets were split into training, validation, and test sets with a ratio of [0.8, 0.1, 0.1]. In our experiments, we set $L = 10$ for the positional encoding step. The model is trained using the Adam optimizer with a learning rate of $1 \times 10^{-3}$ and a weight decay of $1 \times 10^{-5}$. Perception scales for RQ1 are set to [1, 4, 6, 8, 10, 12, 14, 16], and for RQ2 to [1, 2, 4, 8]. The codebook size at each level is fixed at 4096. Experiments are conducted on two Nvidia A-100 GPUs.

\subsection*{Voxel positional encoding}

Prior work demonstrates that models trained directly on raw pixels or spatial coordinates tend to focus on low-frequency content ~\cite{rahaman_spectral_2019} and ignore the high-frequency structures. Positional encoding (PE)~\cite{rahaman_spectral_2019, mildenhall2021nerf} contributes to the recovery of high-frequency details in 3D CV tasks.  To address this, we adopt a sine/cosine PE strategy to represent the high-frequency information on the voxel. The encoding function is defined as $f(v_i)$, where $\mathbf{v}_i$ represents a voxel unit in the density volume, and $L$ denotes the number of encoding levels:

\begin{equation}
f(v_i) = \left( \sin(2^0 \pi v_i), \cos(2^0 \pi v_i), \cdots, \sin(2^{L-1} \pi v_i), \cos(2^{L-1} \pi v_i) \right).
\label{eq:positional_encoding}
\end{equation}

\subsection*{Multi-scale conditional approximations}

The key to VAEs' performance is the quality of latent space approximation via codebook embeddings. Accurately representing the latent feature \( z \) through a well-structured embedding function is central to addressing this challenge. In the VAE quantization process, let \( \hat{z} \) denote the optimal quantized latent vector corresponding to the encoder output \( z \). The goal is to learn a mapping \( z \rightarrow \hat{z} \) through an embedding function \( f_{\theta} \) with parameters \( \theta \). This process is expressed as \( \hat{z} = f_{\theta} \circ z \). Inspired by wavelet approximation theory~\cite{li2024solving,fan_bcr-net_2019}, we can approximate \( f_{\theta} \) using an operator \( \mathcal{A}_{\theta} \). This reformulates the mapping as:

\begin{equation}
\hat{z} \approx \mathcal{A}_{\theta} \circ z
\label{eq:appro_decompose}
\end{equation}

In wavelet theory, the operator \( \mathcal{A}_{\theta} \) can be decomposed into a set of multi-level components $\left[ \mathcal{A}_{\theta}^{(0)}, \cdots, \mathcal{A}_{\theta}^{(l)} \right]$ so that $\mathcal{A}_{\theta} = \sum_{i=0}^{l} \mathcal{A}_{\theta}^{(i)}$.
Under standard linear assumptions, the operator at a given level \( l \) can be expressed in the form:

\begin{equation}
\mathcal{A}_{\theta}^{(l)} = W^{(l)} 
\begin{bmatrix}
D_1^{(l)} & D_2^{(l)} \\
D_3^{(l)} & A^{(l-1)}
\end{bmatrix}
\left( W^{(l)} \right)^\top,
\label{eq:wavelet_level}
\end{equation}

\( W^{(l)} \) denotes the wavelet transformation matrix at level \( l \). \( D_1^{(l)}, D_2^{(l)}, D_3^{(l)} \) are decomposition components. For a detailed theoretical foundation, refer to~\cite{fan_bcr-net_2019, beylkin1991fast}. This hierarchical formulation reveals that the operator at level \( l \) is conditioned on the 
 previous level \( l{-}1 \). This recursive dependency is captured as $\mathcal{A}_{\theta}^{(l+1)} = \mathcal{A}_{\theta}^{(l+1)} \mid \mathcal{A}_{\theta}^{(l)}$. Another wavelet work \cite{li2024solving} demonstrates that incorporating residual structures to fuse conditions enhances signal approximation in nonlinear settings. Inspired by this, we denote the residual conditioning operator as \( \mathcal{M} \). Updating the recursive formulation we have $\mathcal{A}_{\theta}^{(l+1)} = \mathcal{M}(\mathcal{A}_{\theta}^l)$.
\( \mathcal{M} \) represents the residual operation at level \( l \). This allows the approximation operator to capture recursive dependencies across levels, which can be generalized as $\mathcal{A}_{\theta}^{(l+1)} = \mathcal{M} \left( \mathcal{A}_{\theta}^l \mid \mathcal{A}_{\theta}^0, \ldots, \mathcal{A}_{\theta}^{l-1} \right)$. By incorporating this relationship into Equation~\ref{eq:appro_decompose}, the approximation process is formulated as:
\begin{equation}
\hat{z} \approx \sum_{i=0}^{L} 
\mathcal{M} \left( \mathcal{A}_{\theta}^{(i)} \mid \mathcal{A}_{\theta}^{(0)}, \ldots, \mathcal{A}_{\theta}^{(i-1)} \right) \circ z
\label{eq:final_approximation}
\end{equation}

This formulation highlights how recursive conditioning improves approximation quality, offering insight into the improvements of HQ-VAE. Inspired by this, our approach cryo-SWAN explicitly conditions each hierarchical level on the outputs of preceding levels. It differs from HQ-VAE in a key way: each level explicitly conditions on prior outputs through a conditional fusion structure.

\subsection*{Recursive residual quantization}

Besides the multi-level global structure, cryo-SWAN embeds features into the latent space through codebook-based quantization at level $l$. This operation is defined as $\mathcal{A}_{\theta}^{(l)} \circ z = \hat{z}$. Inspired by the recursive perception strategy in VAR-VAE, we decompose the encoder feature map $z$ into a multi-scale representation: $\mathcal{Z} = \{ z^0, z^1, \cdots, z^n \}$. Each perception scale corresponds to a distinct spatial dimension $L_i \times W_i \times z_i$. In contrast to RQ-VAE, each perception level in cryo-SWAN operates with a unique window size and is explicitly conditioned on the preceding level, denoted as $z_i = z_i \mid z_{i-1}$. At scale $l$, the local codebook is defined as $\mathcal{C}^{(l)} = \{ e_1^{(l)}, \cdots, e_k^{(l)} \}$. $e_i^{(l)}$ denotes the $i$-th embedding vector. Quantization operations at level $l$ share $\mathcal{C}^{(l)}$. The optimization objective at level $l$ is given by $\mathcal{C}^{(l)} = \arg\min_j \left\| \mathcal{Z}^{(l)} \mid \mathcal{Z}^{(l-1)} - e_k^{(l)} \right\|_2$. The corresponding loss at level $l$ is defined as in Equation \ref{eq:level_loss}.

\begin{equation}
L_l(\mathcal{A}_\theta^{(l)}) = \sum_{i=1}^{k} \left\| \text{sg} \left[ z_i^{(l)} \mid z_i^{(l-1)} \right] - e_k^{(l)} \right\|^2
\label{eq:level_loss}
\end{equation}


At the global multi-scale level, the total loss is aggregated as $L_{\text{global}} = \sum_{i=1}^{L} L_i \left( A_i \mid A_{i-1} \right)$. This formulation captures the global commitment loss across all levels in the hierarchy.

\subsection*{Cryo-SWAN}
Fig.~\ref{concept} illustrates the architectures of cryo-SWAN. We set the decomposition depth to $L = 2$. The architecture includes two embedding stages (Fig.~\ref{concept}a). The feature maps at each level are quantized in a recursive residual manner (Fig.~\ref{concept}b). Between levels, features are integrated via conditional fusion and propagated to the next stage. Pseudocode in Algorithms~\ref{alg:cryo-swan-compact} and~\ref{alg:ms_vqvae} details the procedures corresponding to Fig.~\ref{concept}a and Fig.~\ref{concept}b. Globally, the total loss consists of the reconstruction loss $\log p(x \mid \hat{z}(x))$ and the sum of commitment losses across all scales, as defined in previous section.

\begin{algorithm}[H]
\caption{cryo-SWAN}
\label{alg:cryo-swan-compact}
\begin{algorithmic}[2]
\State \textbf{Input:} $x \in \mathbb{R}^{B \times 1 \times D \times H \times W}$
\State \textbf{Params:} Scales $K$, $\text{cond.}$
\State \textbf{Output:} $\hat{x}$, $\{\mathcal{L}_k\}_1^K$, $\{I_k\}_1^K$

\Procedure{Forward}{$x$}
    \State \textit{Phase 1: Encoding}
    \State $\text{Enc.} \gets [\,],\, z \gets x$
    \For{$k=1$ \textbf{to} $K$}
        \State $\text{Enc.}\texttt{.push}(\texttt{encoder}_k(z)),\, z \gets \text{Enc.}[k-1]$
    \EndFor
    
    \State \textit{Phase 2: Quant and Decoding}
    \State $\{\mathcal{L}_k\}, \{I_k\} \gets [\,], [\,]$
    \For{k=K to 1}
        \State $z_c \gets \begin{cases}
               \text{Enc.}[k-1] & \text{if } k=K \\
               \texttt{concat}(r_{k+1}, \text{Enc.}[k-1]) & \text{else}
               \end{cases}$
        \State \textbf{if} $\text{cond.} \land k\neq K$: $z_c \gets \texttt{combi\_conv}_k(z_c)$
        \State $\hat{z}_k, I_k, \mathcal{L}_k \gets \texttt{quantizer}_k(\texttt{quant\_conv}_k(z_c))$
        \State $r_k \gets \texttt{decoder}_k(\texttt{post\_quant\_conv}_k(\hat{z}_k))$
        \State $\{\mathcal{L}_k\}\texttt{.append}(\mathcal{L}_k),\, \{I_k\}\texttt{.append}(I_k)$
    \EndFor
    \State \Return $r_1, \{\mathcal{L}_k\}, \{I_k\}$
\EndProcedure
\end{algorithmic}
\end{algorithm}

\begin{algorithm}
\caption{Residual Quantizer}\label{alg:ms_vqvae}
\begin{algorithmic}[1]
\State \textbf{Input:} encoder map $f_{\text{level } i} \in \mathbb{R}^{B \times D \times H \times W}$
\State \textbf{Parameters:} Perception scales $M$
\State \textbf{Output:} $f_{\text{quant}}$, Tokens $R$, Loss $L_i$

\State Initialize: $R_i \gets [\ ]$, $L_i \gets 0$, Codebook $Z \in \mathbb{R}^{V \times C}$

\For{each scale $m \in \{1,\dots,M\}$}
    \State \textnormal{$r_m \gets \arg\min_{j} \|f_m - Z_j\|^2$} 
    \State \textnormal{$R \gets R \cup \{r_m\}$}
    \State \textnormal{$z_m \gets lookup(Z, r_m)$}
    \State \textnormal{$\hat{z}_m \gets \text{Conv3D}(z_m)$} 
    \State \textnormal{$L \gets L + \beta\|f_{\text{quant}} - f_{\text{level } i}\|^2 + \|\hat{z}_m - f_m\|^2$} 
    \State \textnormal{$f_{\text{quant}} \gets f_{\text{quant}} + \hat{z}_m$}
\EndFor

\State{$L_i \gets L_i/M$}
\end{algorithmic}
\end{algorithm}

\subsection*{Determining resolution with Fourier Shell Correlation (FSC)}

To assess the restoration quality of EMDB density maps, we compute the 3D Fourier Shell Correlation (FSC)~\cite{williams_transmission_2009, harauz1986exact}.
In Equation~\ref{eq:fsc}, $k_i$ denotes a voxel in the Fourier space $\mathfrak{F}\{V\}$ corresponding to spatial frequency $k$ (3D Fourier shell) . Computing FSC across the frequency spectrum between reconstructed $V_{\text{pred}}$ and $V_{\text{GT}}$ yields the FSC curve, which reflects the similarity between volumes at different frequencies. This curve captures the signal-to-noise ratio as a function of frequency. The resolution is defined as the highest frequency $k_i$ where the FSC drops below the 0.143 threshold, indicating the last statistically reliable match between the predicted and ground truth volumes~\cite{rosenthal_optimal_2003}.

\begin{equation}
\text{FSC}(k) = \frac{\sum_{k_i \in k} \mathfrak{F}\{V_{\text{pred}}\}(k_i) \cdot \mathfrak{F}^*\{V_{\text{GT}}\}(k_i)}
{\sqrt{\sum_{k_i \in k} \left\| \mathfrak{F}\{V_{\text{pred}}\}(k_i) \right\|^2 \cdot \sum_{k_i \in k} \left\| \mathfrak{F}^*\{V_{\text{GT}}\}(k_i) \right\|^2}}
\label{eq:fsc}
\end{equation}

%% file: sec/2_results.tex
\subsection*{Architecture design for voxel-based 3D shape representation}
\begin{figure}[ht] 
\centering 
\includegraphics[width=0.99\textwidth]{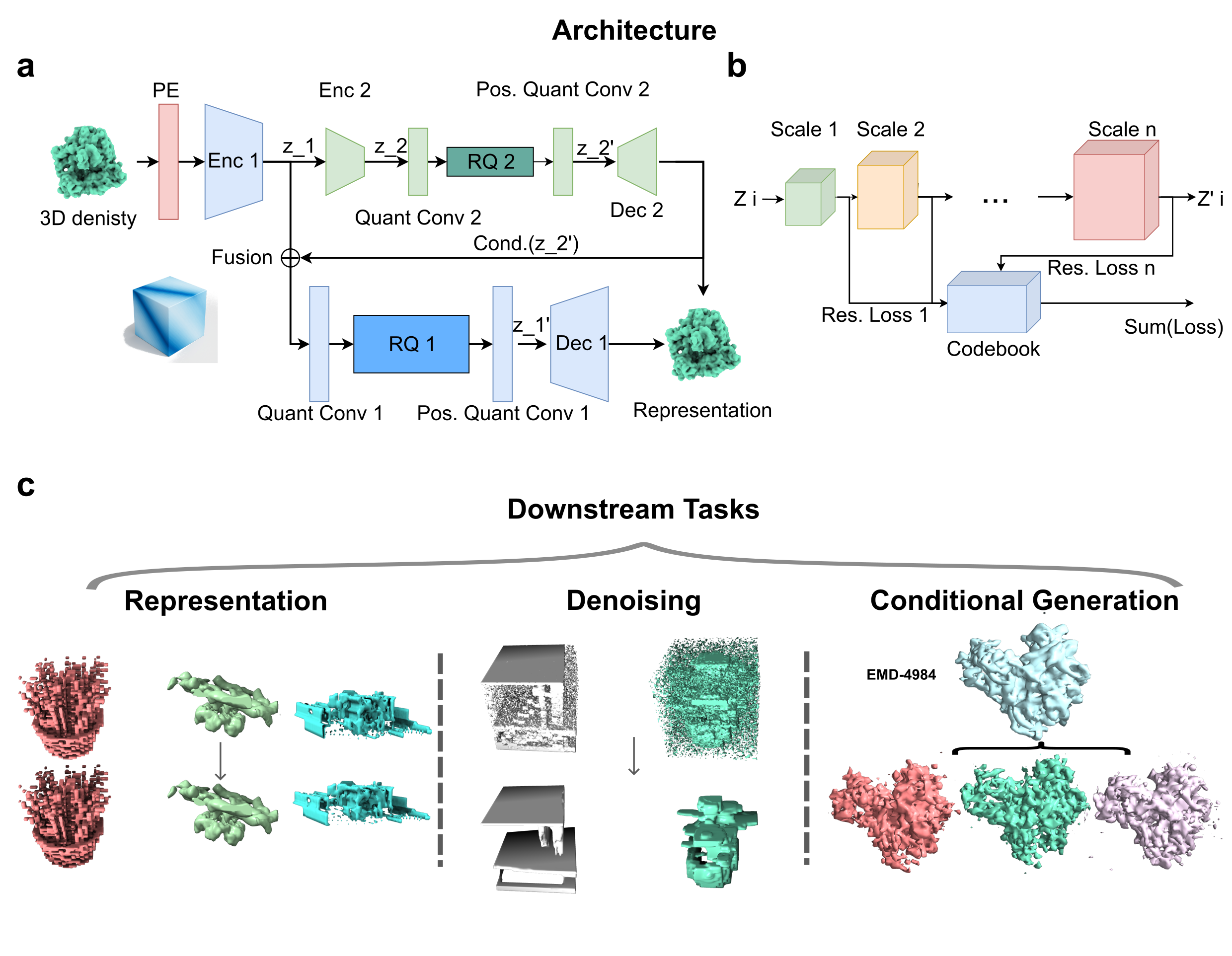}
\caption{\textbf{Principle architecture of Multi-Scale Wavelet-decomposition-inspired Autoencoder Network}. (a) Multi-scale decomposition of the input signal. (b) The residual quantization process. A residual loss is computed during codebook lookup at each scale. (c) Downstream applications of the voxel-based 3D shape representations.} 
\label{concept} 
\end{figure}

To design an architecture capable of learning meaningful representations of shape from voxel-based 3D data, we started with an AE-style structure (Fig.~\ref{concept}a). This was primarily motivated by AEs' abilities to be trained in a self-supervised fashion, allowing them to take advantage of the existing voxel-level datasets. Yet, traditional autoencoders struggle to capture complex geometric properties, such as protein structures, when relying solely on raw density values. Motivated by prior work on multi-level decomposition\cite{li2024solving}, we introduced residual quantization strategies into the latent space and propose a multi-Scale Wavelet-inspired AutoeNcoder cryo-SWAN. Cryo-SWAN iteratively models the latent representation in a multi-scale manner. This enables it to effectively capture both low- and high-frequency geometric information in protein densities.

In this work, specifically, we showcase that the decomposition depth of 2 can provide an efficient density representation up to the used sampling limit of 8 Å. We therefore set the decomposition depth to $L = 2$. The architecture includes two embedding stages. The feature maps at each level are quantized in a recursive residual manner (Fig.~\ref{concept}b). Between levels, features are integrated via conditional fusion and propagated to the next stage (see Methods). Once trained, cryo-SWAN model can be used to perform downstream tasks, such as denoising noisy input volumes or conditional generation of new shapes (Fig.~\ref{concept}c).

\subsection*{Representation learning performance of SWAN on benchmarks}
\begin{figure}[ht] 
\centering 
\includegraphics[width=0.98\textwidth]{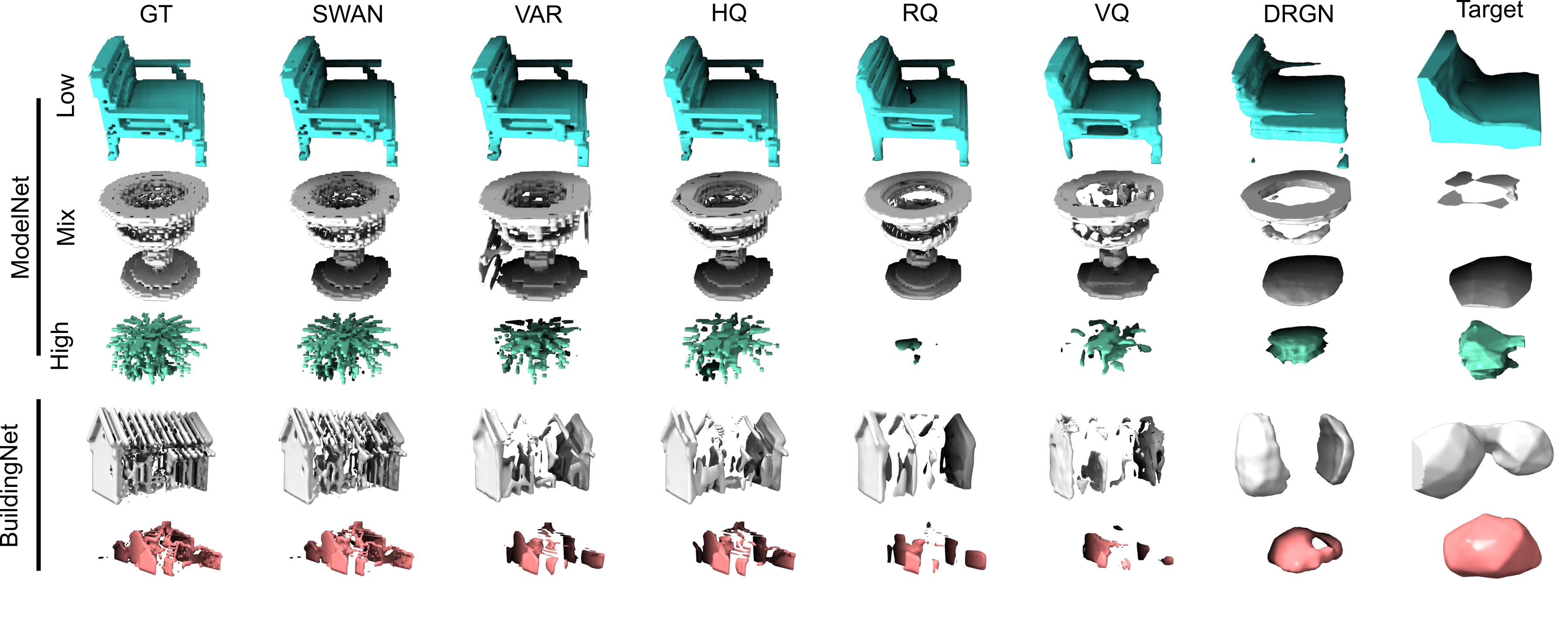}
\caption{\textbf{cryo-SWAN performance on common 3D shape benchmarks}. (a) Performance on three representative objects from BuildingNet dataset. (b) Performance on three representative objects from ModelNet dataset. Here, low, mix, and high denote the frequency components, where mix captures a combination of low- and high-frequency details.} 
\label{fig.generalCV} 
\end{figure}

To evaluate the design of the cryo-SWAN architecture, we conducted representation learning experiments on two benchmarks: ModelNet and BuildingNet. In the first, we selected ModelNet40, which contains 3D objects from 40 categories, ranging from low- to high-frequency shapes. In the second, we used the entire BuildingNet dataset containing structurally complex architectural shapes with rich high-frequency details. To make them comparable with cryo-EM data, we voxelized both datasets into volumetric densities (see Methods). For comparison, we included the SOTA models from computer vision: VAR-VAE, HQ-VAE, RQ-VAE, and the foundational VQ-VAE. Importantly, we focused solely on the VAE components of these models without the generative models. In addition, we evaluated methods tailored to cryo-EM applications, cryo-DRGN\cite{zhong_cryodrgn_2021} and cryo-Target\cite{nasiri2022unsupervised}, which are parts of comprehensive toolkits designed for specific cryo-EM processing tasks. Therefore, to ensure a fair comparison, we evaluate only their autoencoder components, which we call DRGN-VAR and Target-VAE, with respect to representation quality (Fig. \ref{fig.generalCV}). 

To facilitate the qualitative performance comparison on the ModelNet benchmark, we evaluated all models under varying levels of geometric detail. We did this by splitting the objects into low, mix, and high categories, which correspond to predominantly low-frequency, mixed-frequency, and high-frequency geometric information, respectively (Fig. \ref{fig.generalCV}, ModelNet). Results suggest that as the proportion of high-frequency content increases, the representation task becomes progressively more challenging. Specifically, in low-frequency–dominated setting, most methods achieve satisfactory performance. Albeit DRGN-VAE and Target-VAE did not adequately represent the densities in this setting. This limitation likely stems from their reliance on classical VAE backbones, which offer limited representational capacity for complex 3D geometry. As higher-frequency information increased, both RQ-VAE and VQ-VAE lost fine-grained details at the mixed-frequency level. When high-frequency components dominated, cryo-SWAN was the only model that consistently preserved the intricate geometric structures. Results on BuildingNet, which isdominated by high-frequency geometric details, suggested the same trends (Fig. \ref{fig.generalCV}, BuildingNet). Cryo-SWAN effectively captured such frequency information and demonstratedsuperior performance over all competing methods.

Next, to quantitatively evaluate the performance, we used four standard metrics: MSE (0.007), PSNR (21.726), IoU (0.993), and F1 score (0.996), shown in Tab. \ref{Table.Evaluate}. Results suggest that across general computer vision (CV) benchmarks, our model consistently outperformed all competing methods on all metrics, demonstrating a clear performance advantage.

\begin{table*}[ht]
  \centering
  \begin{adjustbox}{width=0.98\textwidth,center}
    \begin{tabular}{lccccccccccccc}
      \toprule
      Model & \multicolumn{4}{c}{ModelNet} & \multicolumn{4}{c}{BuildingNet} & \multicolumn{5}{c}{ProteinNet3D} \\
      \cmidrule(lr){2-5} \cmidrule(lr){6-9} \cmidrule(lr){10-14}
      & MSE$\downarrow$ & PSNR$\uparrow$ & IoU$\uparrow$ & F1$\uparrow$ 
      & MSE$\downarrow$ & PSNR$\uparrow$ & IoU$\uparrow$ & F1$\uparrow$
      & MSE$\downarrow$ & PSNR$\uparrow$ & IoU$\uparrow$ & F1$\uparrow$ & FSC (0.5)$\downarrow$ \\
      \midrule
      SWAN & \textbf{0.007} & \textbf{21.726} & \textbf{0.993} & \textbf{0.996}
           & \textbf{0.003} & \textbf{25.585} & \textbf{0.839} & \textbf{0.912}
           & \textbf{0.003} & \textbf{27.063} & \textbf{0.672} & \textbf{0.799} & \textbf{9.10} \\
      VAR  & 0.025 & 17.508 & 0.836 & 0.904
           & 0.006 & 22.736 & 0.616 & 0.761
           & 0.006 & 22.790 & \underline{0.400} & \underline{0.558} & \underline{14.01} \\
      HQ   & \underline{0.013} & \underline{19.612} & \underline{0.924} & \underline{0.958}
           & \underline{0.005} & \underline{23.588} & \underline{0.672} & \underline{0.802}
           & 0.007 & 22.558 & 0.317 & 0.466 & 18.20 \\
      RQ   & 0.027 & 16.326 & 0.790 & 0.875
           & 0.008 & 21.886 & 0.476 & 0.642
           & 0.008 & 21.990 & 0.267 & 0.409 & 24.08 \\
      VQ   & 0.044 & 13.931 & 0.726 & 0.834
           & 0.025 & 16.374 & 0.327 & 0.489
           & \underline{0.005} & \underline{24.107} & 0.248 & 0.388 & 23.49 \\
      DRGN & 0.055 & 13.201 & 0.518 & 0.660
           & 0.025 & 16.983 & 0.207 & 0.341
           & 0.012 & 20.210 & 0.182 & 0.300 & 55.75 \\
      Target & 0.119 & 9.740 & 0.312 & 0.458
             & 0.049 & 13.480 & 0.147 & 0.253
             & 0.011 & 20.394 & 0.140 & 0.238 & 77.82 \\
      \bottomrule
    \end{tabular}
  \end{adjustbox}
  \caption{Evaluation of candidates on three datasets with PSNR, MSE, IoU, F1-score, and FSC resolution (only for Protein3DNet). Bold items indicate the best performance, underlined values indicate the second best.}
  \label{Table.Evaluate}
\end{table*}

\subsection*{SWAN performance on 3D molecular shapes}
To assess cryo-SWAN on experimentally derived molecular data, we curated ProteinNet3D, comprising over 24,000 cryo-EM density volumes retrieved from EMDB. All maps were resampled to a uniform isotropic voxel spacing of 4 Å, resulting in a Nyquist limit of 8~\AA~\cite{noauthor_nyquist-shannon_nodate}. At this sampling, protein densities retain substantial structural complexity while exhibiting heterogeneous noise levels and diverse geometric features across multiple spatial scales, which presents a great challenge for representation learning algorithms due to highly complex geometries~\cite{harris_bound_1981}. This makes ProteinNet3D a stringent benchmark for volumetric representation learning.

\begin{figure}[ht] 
\centering 
\includegraphics[width=0.85\textwidth]{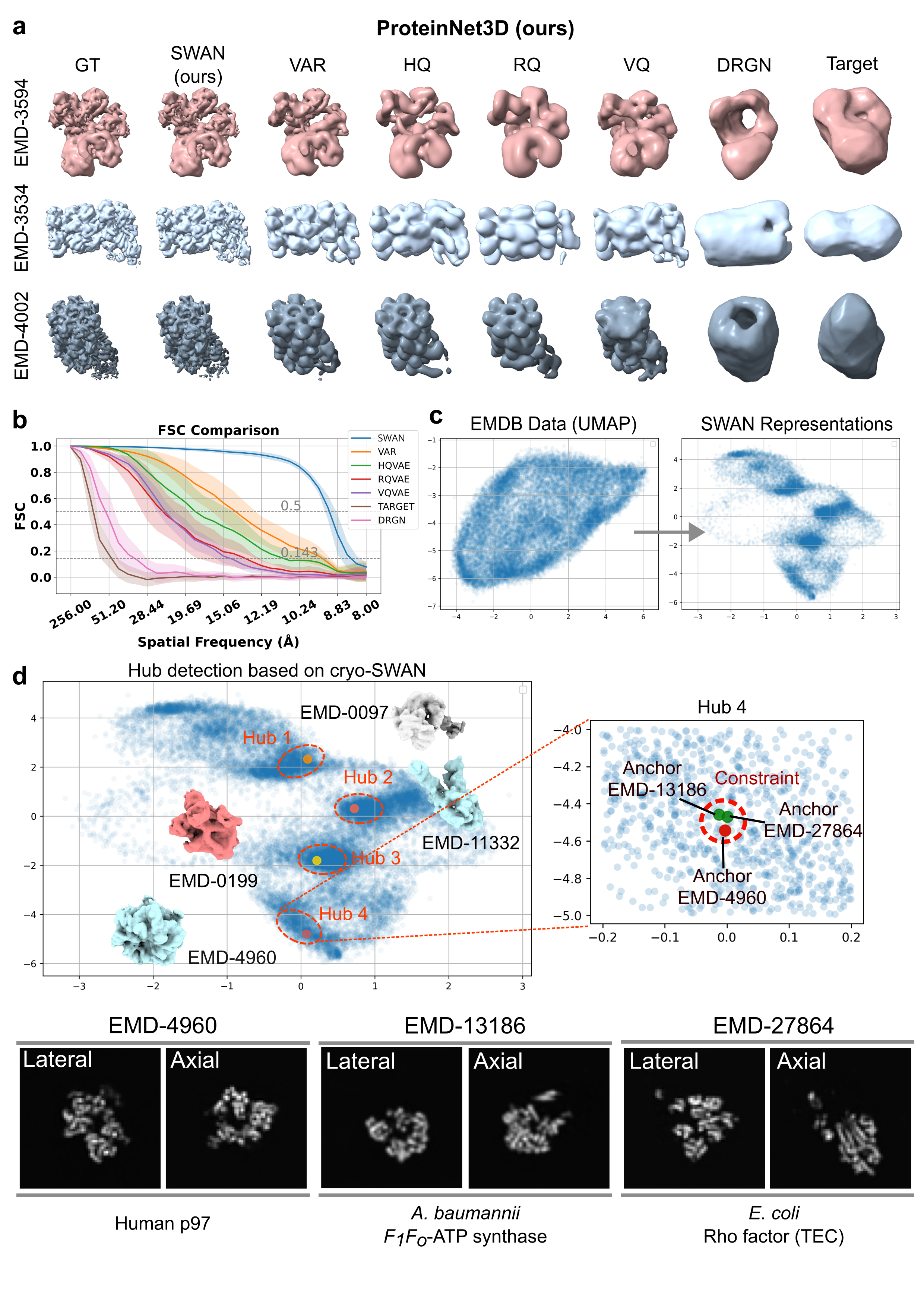}
\caption{\textbf{Representation learning on the ProteinNet3D and explorations in the latent space.} 
(a) The representation learning on the ProteinNet3D. (b) The FSC evaluation of all the representations. (c) Dimensionality reduction comparison: Cryo-SWAN latent vectors vs. original EMDB volumes. (d) UMAP projection of latent vectors reveals distinct hubs reflecting structural similarity. Molecules with similar 3D structures at 8~\AA\ resolution (siblings).} 
\label{Fig.proteinNet3D} 
\end{figure}

Representative reconstructions produced by cryo-SWAN and competing methods are shown in Fig.~\ref{Fig.proteinNet3D}a. Remarkably, qualitative comparisons indicate that DRGN-VAE and Target-VAE predominantly reconstruct low-frequency structural components, with limited recovery of finer geometric features. VAR-VAE and HQ-VAE better capture intermediate detail, yet visibly attenuate high-frequency density variations. In contrast, cryo-SWAN preserves fine-grained structural elements while maintaining coherent global shape.

We next evaluated the performance of all models quantitatively (Tab. \ref{Table.Evaluate}). Results suggest that the cryo-SWAN outperforms other models on all the metrics: MSE (0.003), PSNR (27.063), IoU (0.672), and F1 (0.799). To further interrogate the performance of the models with respect to a commonly used cryo-EM measure, we evaluated the Fourier Shell Correlation (FSC) between the input volumes and reconstructions (Fig. \ref{Fig.proteinNet3D}). Results suggested that the cryo-SWAN achieved a formidable resolution of 9.10 \text{\AA} (with a cut-off of 0.5). Altogether, these results suggest that our architecture (cryo-SWAN) consistently outperforms other models in reconstructing complex 3D densities as protein structures across all the metrics, which can be attributed to the architecture's ability to learn higher-quality representations.

To further interrogate this point, we performed a dimensionality reduction analysis of the SWAN representations using UMAP \cite{McInnes2018} (Fig. \ref{Fig.proteinNet3D}c). Specifically, we compared UMAP visualizations of the original EMDB data points with their latent representations from the cryo-SWAN model. Evidently, the cryo-SWAN embeddings exhibit richer and clearer structural organization of 3D protein densities. Visualisation suggests that certain data points are positioned more closely, likely reflecting their higher geometric similarity. Further analysis of such points in Fig. \ref{Fig.proteinNet3D}d reveals that certain proteins exhibit strong geometric similarity, which we refer to as hubs. By zooming into a representative hub, we identify an anchor protein density and retrieve other geometrically similar densities within a defined neighborhood. For example, EMD-4960 (human p97/VCP), EMD-13186 (\textit{A. baumannii} $F_1F_o$-ATP synthase), and EMD-27864 (\textit{E. coli} Rho factor) occupy a neighboring region of the latent manifold. At the ~8 Å resolution considered here, these densities share common geometric features observable at the envelope level, including ring-like features with a rotational symmetry around the central axis, a resolvable internal cavity, and similar size and aspect ratio. We present more examples at the supplementary Fig.\ref{twins}. These structurally similar densities highlight shared geometric characteristics and may provide insights into their underlying biological functions.  

Importantly, the observed latent space neighborhoods reflect similarity in volumetric geometry rather than sequence identity or biochemical function. These results suggest that cryo-SWAN embeddings capture multi-level structural organization at the level of resolvable density features, indicating that the latent space encodes structural similarities within cryo-EM data. Together, these results demonstrate that cryo-SWAN not only improves reconstruction fidelity on complex molecular volumes but also learns latent representations that encode meaningful geometric structure within cryo-EM data.

\subsection*{Ablation study}
\begin{figure}[ht] 
\centering 
\includegraphics[width=0.85\textwidth]{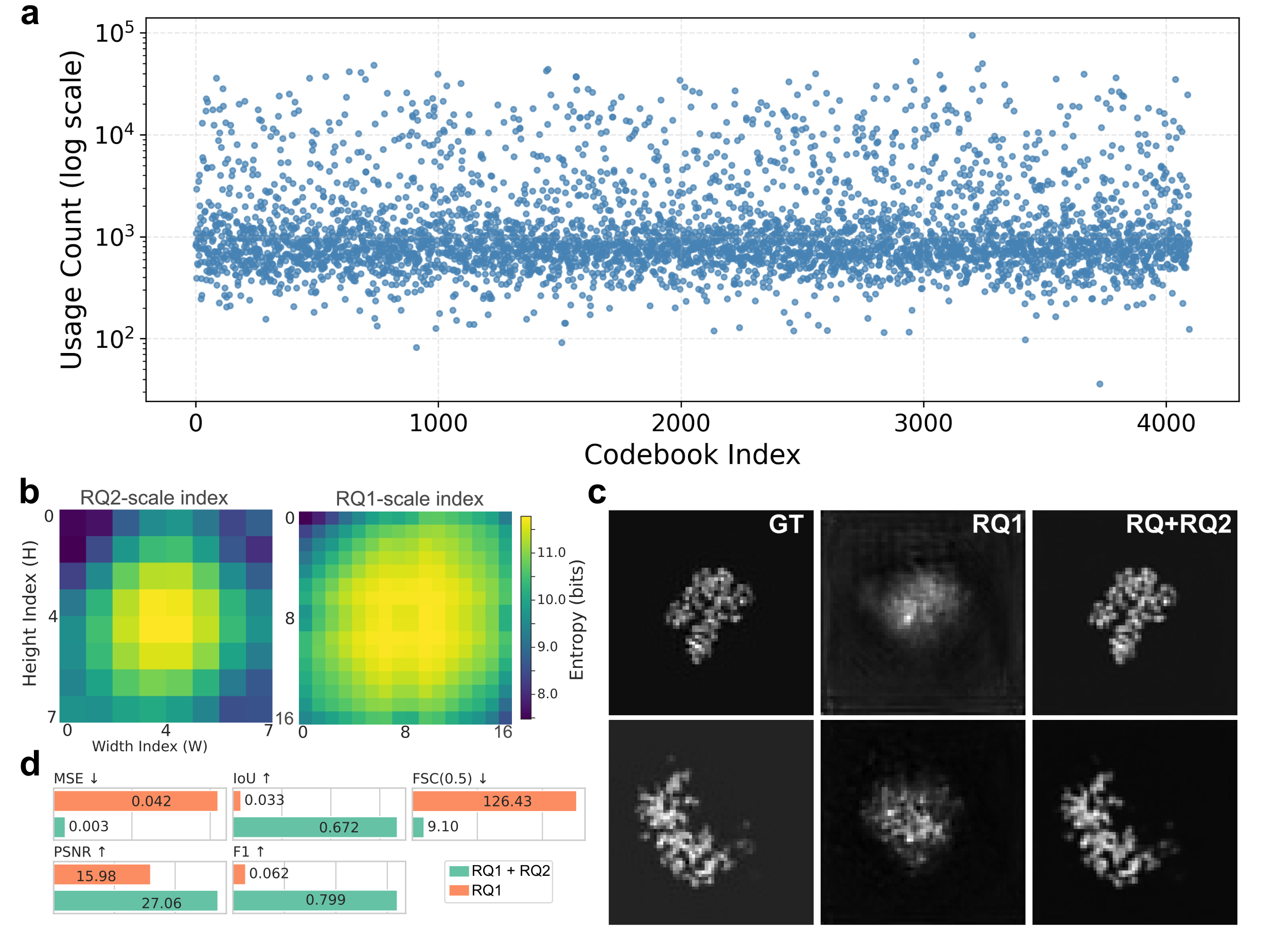}
\caption{\textbf{Ablation study for codebook collapsing and multi-scale representations.}
(a) The calls for all the codebook entries. (b) The latent code representation for both scales (coarse and fine). (c) The molecular density volume: original (GT), at a certain representation scale (RQ1) and at a full representation (RQ1+RQ2). (d) Evaluation of representation performance at a certain representation scale (RQ1) compared to the full representation (RQ1+RQ2).
}
\label{ablation} 
\end{figure}

To verify that the performance gain of cryo-SWAN results from our design choices, we performed an ablation study. A key challenge in VQ-VAE-based models involves codebook collapse~\cite{takida_hq-vae_2024, kossale_mode_2022}, where only a small subset of the codebook indices is used during embedding. Such a collapse often leads to inefficient codebook utilization and degraded embedding quality. An effective embedding strategy should promote both diverse index usage and semantically meaningful patterns. Therefore, to assess the codebook usage, we visualized the lateral view of the latent embedding indices from residual quantization level (RQ1; see Fig.~\ref{concept}a,b) on ProteinNet3D (Fig.\ref{ablation}a). We examined quantization behavior across perception scales (4, 8, 16). Results suggest that the codebook index is well-utilized throughout with no obvious signs of collapse.

We further zoomed in on specific patches (Fig.\ref{ablation}b) across the perception scales RQ1 and RQ2 (see Fig.~\ref{concept}a,b for architecture reference). The results reveal a clear coarse-to-fine quantization pattern: lower scales use broader index assignments, while higher scales introduce localized details. Transitions across scales remain smooth and coherent, suggesting that the model recursively builds hierarchical features in a structured and consistent manner.

Finally, we examined the molecular shape reconstructions performance from RQ1 in comparison to the full representation (Fig.\ref{ablation}c) visually and quantitatively (Fig.\ref{ablation}d). The single-scale variant exhibits noticeably degraded reconstruction quality, particularly in high-frequency regions, and performs significantly worse across evaluation metrics. These findings suggest that the single-scale performance (RQ1) is significantly worse than the multi-scale (RQ1+RQ2), justifying our architectural design choices.

\subsection*{Downstream tasks of cryo-SWAN}
\begin{figure}[ht!] 
\centering 
\includegraphics[width=0.99\textwidth]{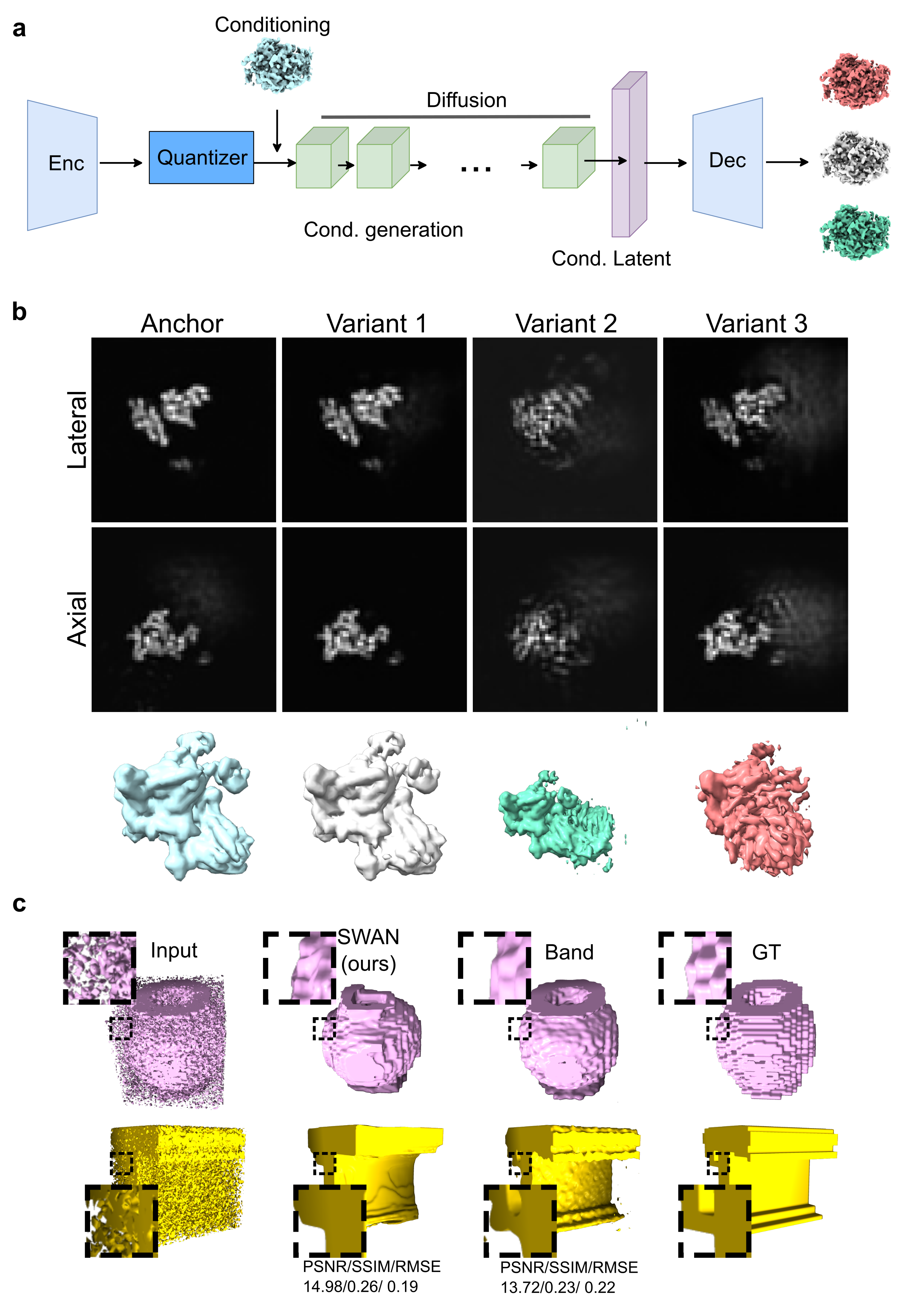}
\caption{\textbf{Downstream applications of cryo-SWAN including shape denoising and conditional molecular shape generation.}
(a) The conditional generation from diffusion models based on cryo-SWAN representations. (b) Examples of conditional generations. The variants are similar in the geometric perspective to the real protein densities in EMDB (anchor point). (c) The unsupervised denoising based on the representation learning for 3D densities from ModelNet.
}
\label{appl} 
\end{figure}
Finally, to demonstrate how cryo-SWAN high-quality representations can be used for downstream applications, we explore two tasks: conditional generation of 3D molecular structures and unsupervised 3D denoising.

In the first task, the high-quality representations cryo-SWAN learned can be employed to generate similar shapes in high resolution (pixel size as 4 \text{\AA} in this work), given an input shape as a condition. Albeit indirectly, this can serve a number of applications from \textit{de novo} molecular design to synthetic data generation. To accomplish this, we used the denoising diffusion probabilistic model (DDPM) \cite{ho_denoising_2020} to learn the distribution of latent vectors from the cryo-SWAN quantization (Fig.~\ref{appl}). During generation, we condition the molecules' latent vectors on DDPM, enabling the synthesis of structurally similar variants (Fig.~\ref{appl}a). These changed latent vectors are decoded by the cryo-SWAN decoder to produce synthetic 3D molecular density maps (Fig.~\ref{appl}b). Results suggest that the generated densities preserve key structural features of the anchor. For more examples, please refer to supplementary Fig.\ref{gens}.

Next, we explored the downstream task of unsupervised 3D denoising (Fig.~\ref{appl}c). To simulate the reconstruction of the noisy input we first added artificial Gaussian noise (SNR=0.5) to the shapes from the ModelNet40 benchmark and then restored it using a band-pass filter (low-0.25, high-0.01) or cryo-SWAN (see Methods). By projecting latent codes of noisy inputs onto the latent space of clean densities, our approach achieves effective denoising while preserving high-frequency information. Results suggest that cryo-SWAN was able to better preserve the high-frequency (fine-grained) 3D structural details of the 3D shape compared to band-pass (Fig.~\ref{appl}c, zoomed-in boxes). Furthermore, our approach outperformed the baseline with respect to PSNR (14.98), SSIM (0.26), and RMSE (0.19) metrics on the test dataset. These results indicate that the learned latent representations can act as a structured prior over valid 3D shapes, enabling restoration beyond simple frequency-domain filtering.

%% file: sec/3_discussion.tex
Learning representations directly from the voxelized shapes is an under-researched avenue, yet it bears promise for creating advanced algorithms in the realm of biomedical imaging. It opens an avenue for handling data from modalities like cryo-ET directly on the averaged densities, allowing us to learn high-quality representations of molecular shapes.

In this paper, we propose addressing this challenge by introducing cryo-SWAN – a compute-efficient architecture inspired by wavelet decomposition. Cryo-SWAN performs conditional, coarse-to-fine residual quantization at multiple scales, allowing to capture both global shape and fine structural details. We evaluated cryo-SWAN on three datasets: ModelNet and BuildingNet (voxelized), and ProteinNet3D  – a large-scale cryo-EM dataset we curated using EMDB\cite{the_wwpdb_consortium_emdbelectron_2024} as a source. We demonstrate that cryo-SWAN outperformed SOTA pixel/voxel-based VAE baselines (HQ-VAE, VAR-VAE, RQ-VAE, and VQ-VAE) on metrics: MSE, PSNR, IoU, and F1-score across all datasets. 

These improvements arise from two core design principles. First, the coarse-to-fine conditional architecture explicitly distributes representational capacity: the coarse level captures overall geometry, while the fine level focuses on localized high-resolution features. This hierarchical separation mitigates the blurring and detail loss common to single-scale latent models. Second, recursive residual quantization enforces structured latent approximations that promote diverse and semantically meaningful codebook utilization, reducing collapse phenomena often observed in vector-quantized models. Ablation experiments confirm that this multi-scale design is essential to the observed performance gains.

Furthermore, we explored the quality of the learned representations through UMAP dimensionality reduction. Our analysis revealed clear data points aggregates (hubs). Molecules in these hubs shared similar structural characteristics. It is tempting to speculate that cryo-SWAN and future approaches could be used for structural mining, linking molecular geometry with putative links to biological function. Finally, we explored conditional shape generation – a downstream task enabled by the high-quality representations learned by cryo-SWAN. We implemented a latent diffusion pipeline based on cryo-SWAN and DDPM\cite{ho_denoising_2020}. By conditioning the DDPM on an anchor protein, we generated synthetic density volumes with similar structural characteristics, enabling controllable and realistic molecular variation.

Notably, some limitations persist due to the scope of this work. Firstly, we explored only a two-scale decomposition. However, further scales could be possible, albeit likely with diminishing returns. Secondly, cross-scale codebook sharing remains an open question, as does using different codebook sizes per scale. We aim to explore these questions in future.

Practically, cryo-SWAN may enable the construction of data-driven molecular shape priors directly from experimental density repositories. Such priors may improve reconstructions, support realistic data augmentation, and reduce reliance on models when they are incomplete or unavailable. Potential extensions include molecular identification through latent-space, restoration of tomographic "missing-wedge" artifacts \cite{wiedemann2024deep}, and orientation estimation of molecules for subtomogram averaging \cite{LEIGH2019217}.

Taken together, cryo-SWAN provides a general-purpose framework for learning 3D representations of volumetric densities. Our work opens several promising avenues for representation learning in structural biology and possibly other fields relying on 3D biomedical imaging. Looking forward, Cryo-SWAN’s architecture could be extended with multi-modal data by incorporating a sequence encoder~\cite{evolutionary-scale, proteinbert}. Coupled with its high-resolution representation capability, this could further facilitate the rational design of molecular sequences with defined structures.

%% file: sec/5_acknowlegements.tex
This work was partially funded by the Center for Advanced Systems Understanding (CASUS), which is financed by Germany’s Federal Ministry of Research, Technology and Space (BMFTR) and by the Saxon Ministry for Science, Culture, and Tourism (SMWK) with tax funds based on the budget approved by the Saxon State Parliament. The work was supported by the Helmholtz Imaging IVF grant CryoFocal. MK is supported by the Heisenberg award from the DFG (KU 3222/2-1), and funding from the Helmholtz Association. AY is supported by the Helmholtz Association Initiative and Networking Fund in the frame of Helmholtz AI as well as by the Helmholtz Foundation Model Initiative within the project “PROFOUND”. The authors thank HelmholtzAI (grant tomoCAT). The authors gratefully acknowledge the Gauss Centre for Supercomputing e.V. (www.gauss-centre.eu) for funding this project by providing computing time through the John von Neumann Institute for Computing (NIC) on the GCS Supercomputer JUWELS at Jülich Supercomputing Centre (JSC).

%% file: sec/6_suppl.tex
\setcounter{figure}{0} 
\renewcommand{\thefigure}{S\arabic{figure}} 

\subsection*{ProteinNet3D}

\begin{figure}[H] 
\centering 
\includegraphics[width=0.98\textwidth]{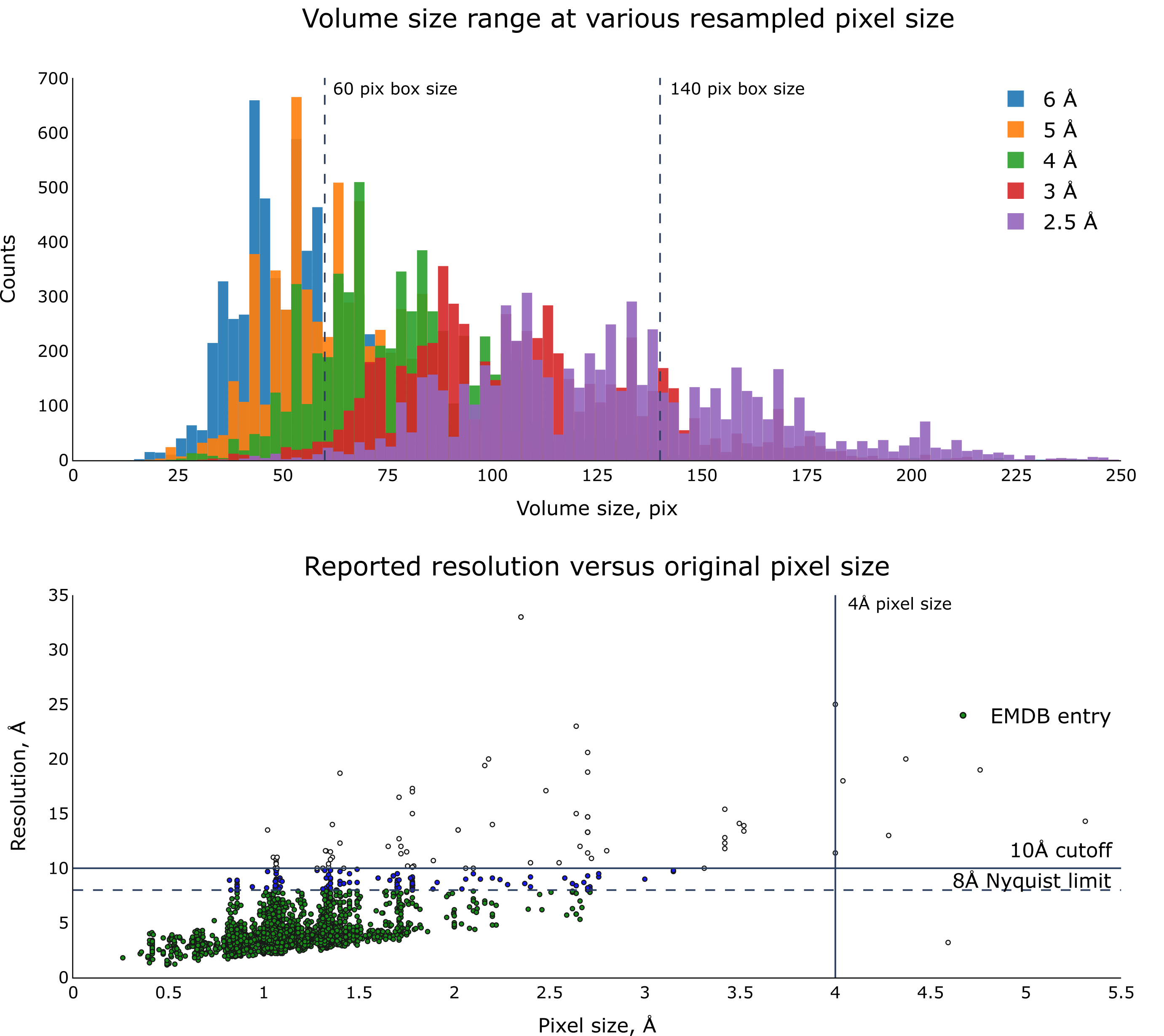}
\caption{ \textbf{Optimal choice of the pixel size and the volume size.} (top) Volume size distribution at various resampled pixel size. (bottom) Volumes distribution respective to the originally reported resolution and the pixel size: ProteinNet3D volumes, preserving original resolution after resampling (blue) and degraded to $8\,\text{\AA}$ after resampling due to the Nyquist limit (green), as well as filtered out volumes due to the cutoffs applied (white). }
\label{metadata_apix} 
\end{figure}

\begin{figure}[H] 
\centering 
\includegraphics[width=0.8\textwidth]{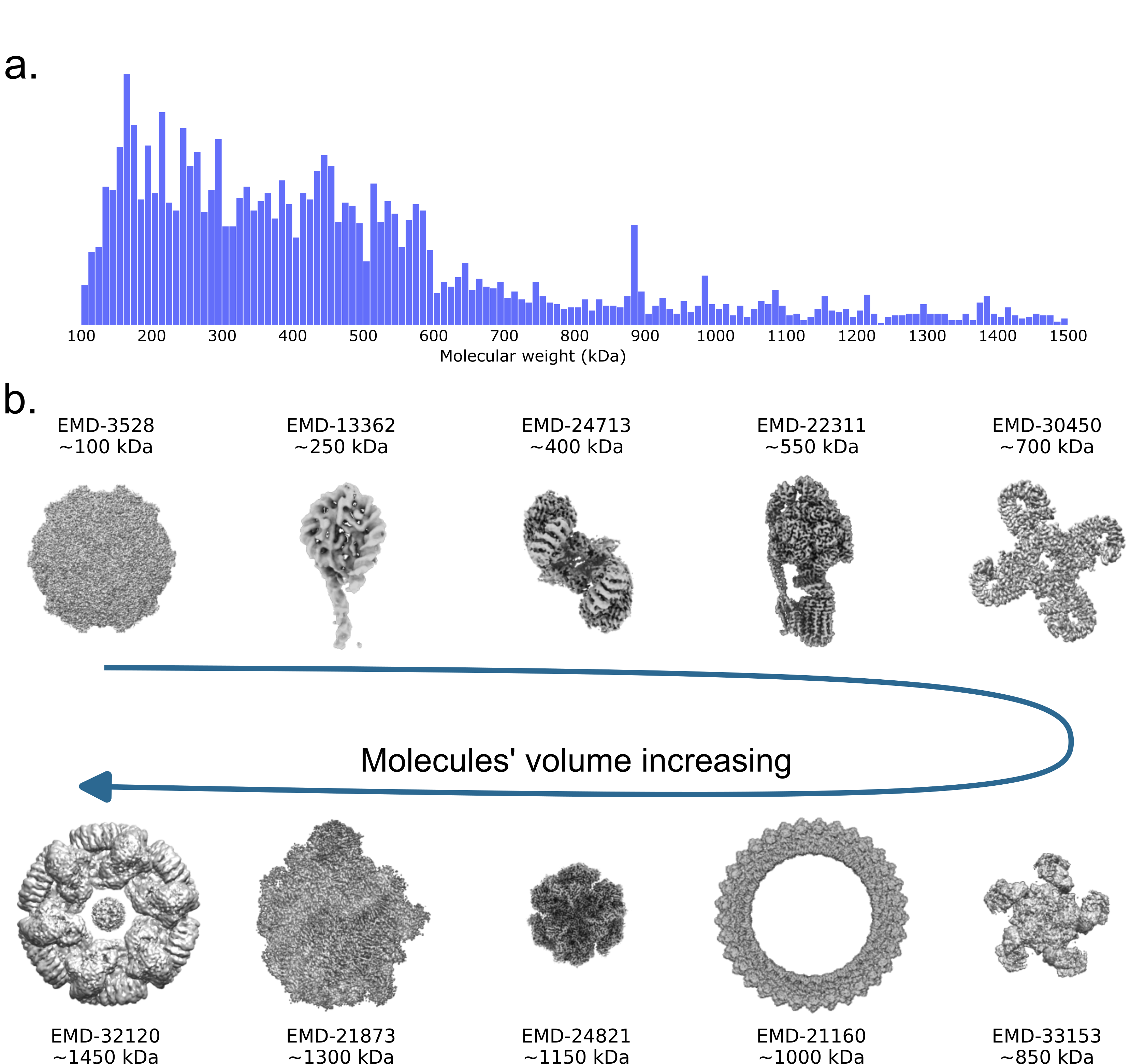}
\caption{ \textbf{ProteinNet3D diversity: molecular weights range.} (a) Molecular weights distribution. (b) Example molecular volume views }
\label{metadata_mw} 
\end{figure}

\begin{figure}[H] 
\centering 
\includegraphics[width=0.8\textwidth]{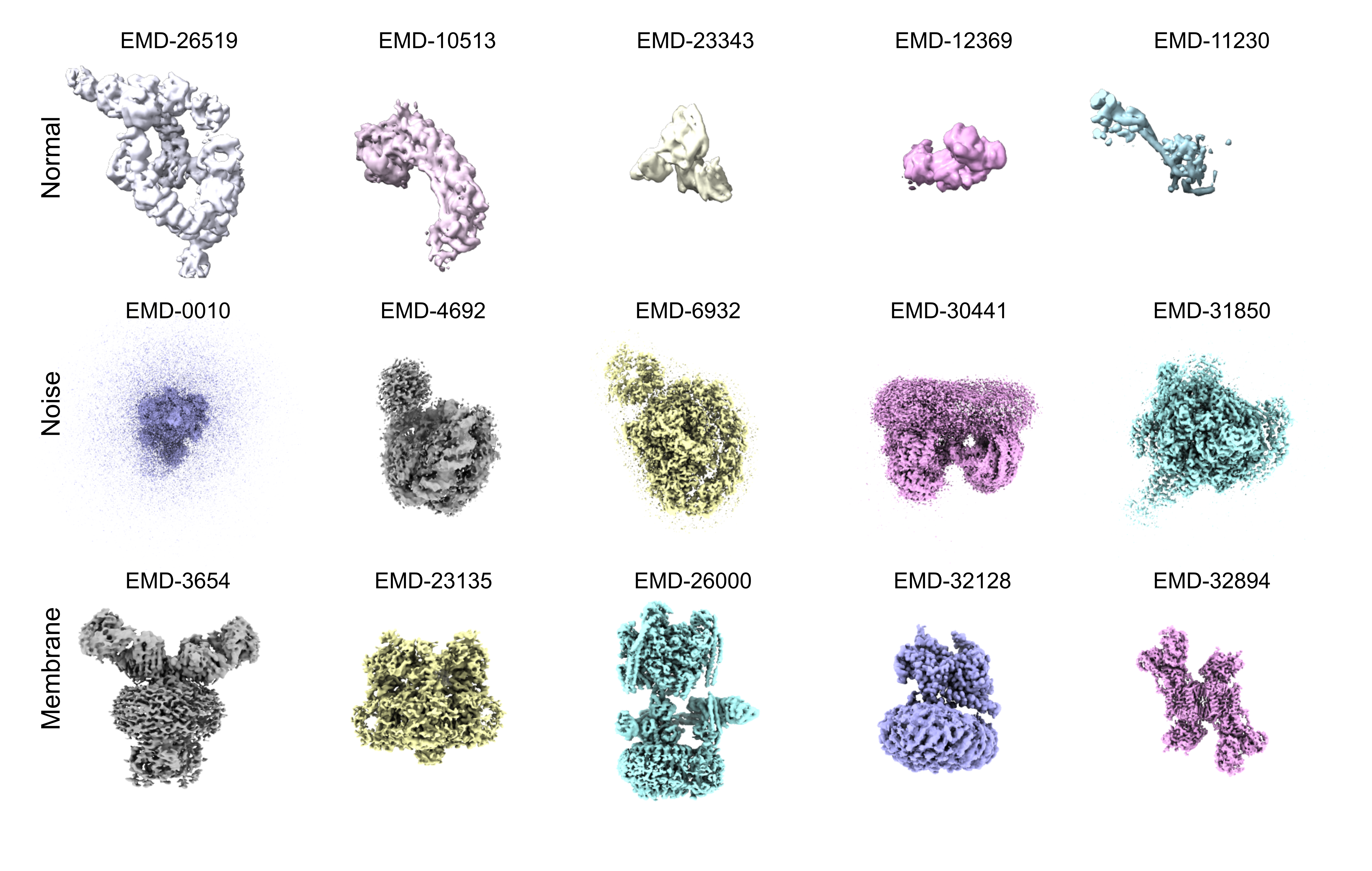}
\caption{ \textbf{ProteinNet3D complexity: membrane-related signal and experimental noise  presence.} 
(top) Normal protein densities in ProteinNet. (middle) ProteinNet example volumes containing membrane-related signal. (bottom) ProteinNet example volumes containing experimental noise }
\label{complexity} 
\end{figure}

Here we offer additional details on the preparation and contents of the ProteinNet3D dataset, including the optimal volume and pixel size choice (Supplementary Fig.~\ref{metadata_apix}), the diversity of the molecular weights (Suppl. Fig.~\ref{metadata_mw}), and the variability of background scenarios (Supplementary Fig.~\ref{complexity}). 

\subsection*{Applications}

We present additional examples of geometry-similarity detection in Supplementary Fig.~\ref{twins}. Leveraging Cryo-SWAN, we identify proteins with similar structural geometry at 8~\AA\ resolution. These findings may offer insights into the relationship between molecular shape and biological function.

\begin{figure}[H] 
\centering 
\includegraphics[width=0.95\textwidth]{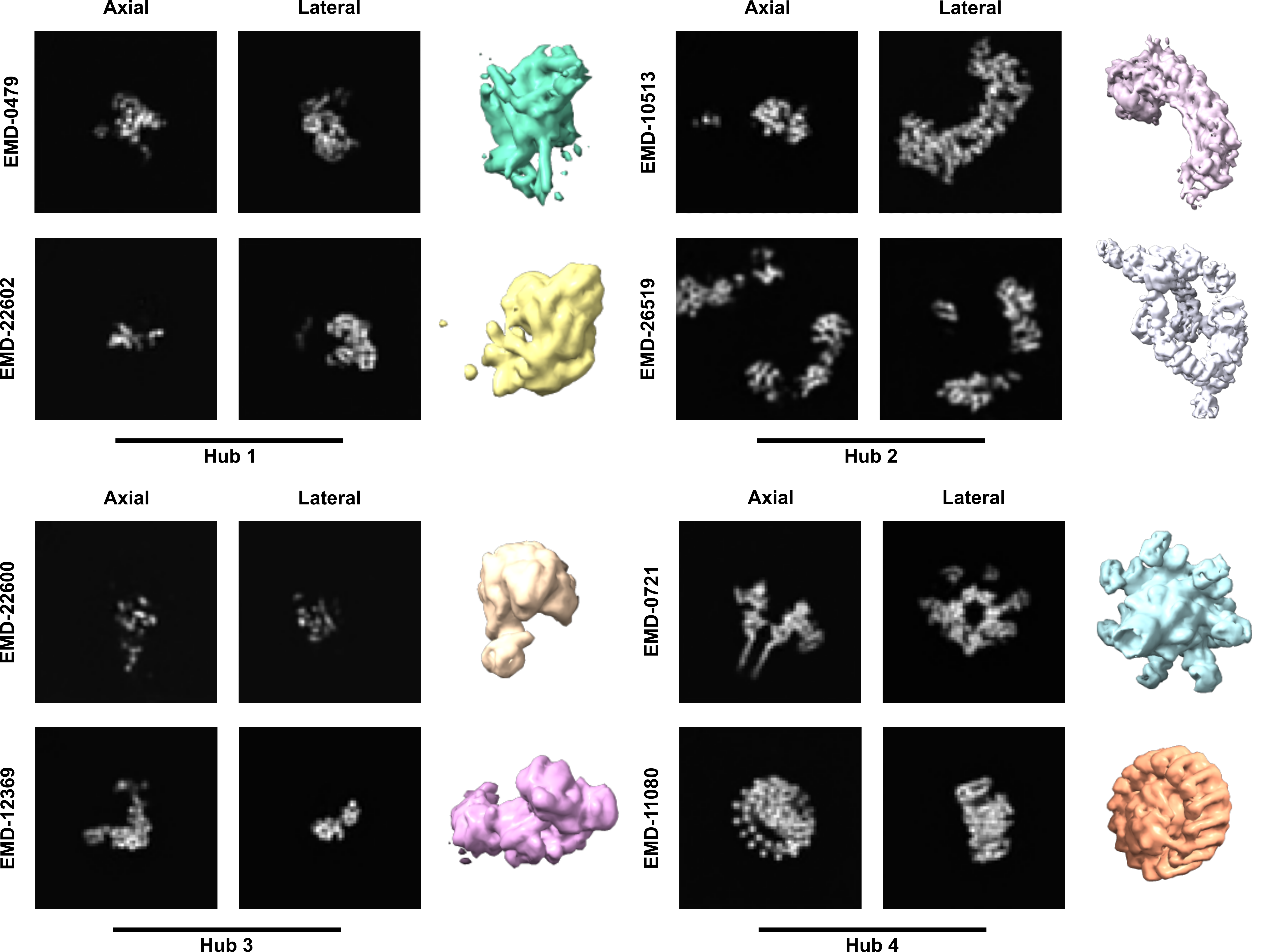}
\caption{\textbf{Molecular shapes similarity within hubs identified with Cryo-SWAN.}} 
\label{twins} 
\end{figure}

Additional generation examples are shown in Supplementary Fig.~\ref{gens}. Using the latent representation of a template protein (anchor), Cryo-SWAN enables the generation of structurally similar 3D densities. Note that these results were produced using a basic 3D U-Net DDPM, leading to some variation and scatter in the outputs. Higher-quality generation could be achieved by incorporating more advanced generative models, such as GANs, transformers, and Stable Diffusion.

\begin{figure}[H] 
\centering 
\includegraphics[width=0.95\textwidth]{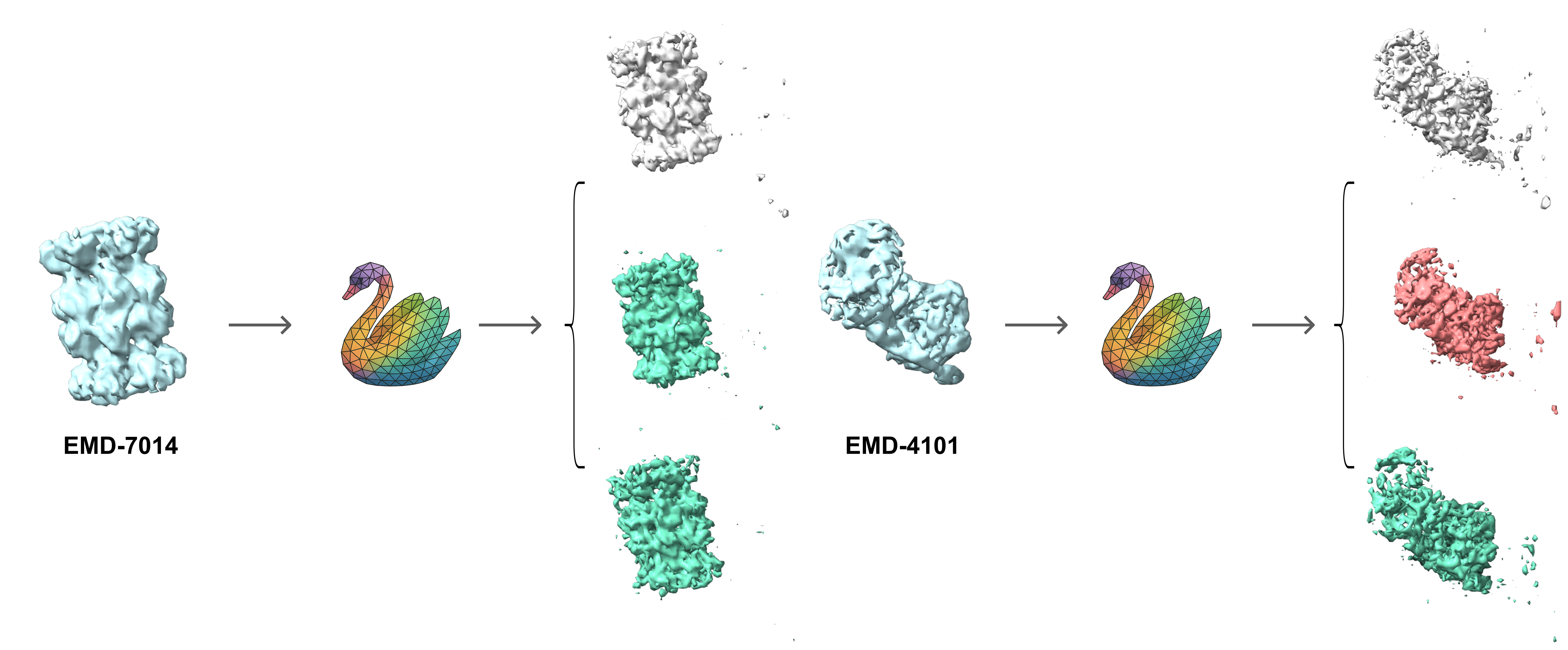}
\caption{\textbf{The latent diffusion generation based on cryo-SWAN representation.}} 
\label{gens} 
\end{figure}